\documentclass[twoside,twocolumn,9pt]{article}
\usepackage{extsizes}
\usepackage[super,sort&compress,comma]{natbib} 
\usepackage[version=3]{mhchem}
\usepackage[left=1.5cm, right=1.5cm, top=1.785cm, bottom=2.0cm]{geometry}
\usepackage{balance}
\usepackage{mathptmx}
\usepackage{sectsty}
\usepackage{graphicx} 
\usepackage{lastpage}
\usepackage[format=plain,justification=justified,singlelinecheck=false,font={stretch=1.125,small,sf},labelfont=bf,labelsep=space]{caption}
\usepackage{float}
\usepackage{fancyhdr}
\usepackage{fnpos}
\usepackage[english]{babel}
\usepackage{array}
\usepackage{droidsans}
\usepackage{charter}
\usepackage[usenames,dvipsnames]{xcolor}
\usepackage{setspace}
\usepackage[compact]{titlesec}
\usepackage{hyperref}
\usepackage[utf8]{inputenc}

\usepackage{epstopdf}
\usepackage{xcolor}

\definecolor{cream}{RGB}{222,217,201}
\usepackage{amsmath,amssymb}

\usepackage{authblk}

\begin{document}

\title{Phase separation in soft repulsive polymer mixtures: Foundation and implication for chromatin organization}
\author[1]{Naoki Iso}
\author[1]{Yuki Norizoe}
\author[1]{Takahiro Sakaue$^{\ast}$}
\affil[1]{Department of Physical Sciences, Aoyama Gakuin University}
\date{}
\maketitle
\begin{abstract}
Given a wide range of length scales, the analysis of polymer systems often requires coarse-graining, for which various levels of description may be possible depending on the phenomenon under consideration. Here, we provide a super-coarse grained description, where polymers are represented as a succession of mesosopic soft beads which are allowed to overlap with others.
We then investigate  the phase separation behaviors in mixture of such homopolymers based on a mean-field theory, and discuss universal aspects of the miscibility phase diagram in comparison with the numerical simulation. We also discuss an extension of our analysis to the mixtures involving random copolymers, which might be interesting in the context of chromatin organization in cellular nucleus.
\end{abstract}

\footnotetext[0]{\textit{$^{*}$~Department of Physical Sciences, Aoyama Gakuin University, 5-10-1 Fuchinobe, Chuo-ku, Sagamihara, Japan. E-mail:sakaue@phys.aoyama.ac.jp}}

\section{Introduction}
Phase separations in polymer solution and blend have a long history of research due to its importance in fundamental science as well as industrial applications~\cite{deGennes_scaling_concept, doi1995introduction, BATES_1991, Witte_1996, Yin_2022, Choon_2022, Sakuta_2020}. Recently, its pivotal role in the field of biophysics has been recognized as a basic mechanism to organize various cellular and nuclear bodies~\cite{Brangwynne_2009, Chen_2019, Esposito_2022, Yanagisawa_2022, Falk_2019, Riback_2020, Deviri_2021, Yamazaki_2022, Yamamoto_2023, Lennart_2021, Michieletto_2019, Koyama_2024}.
Here, given the complexity in biological systems, standard approaches such as the Flory-Huggins theory to analyze the phase separation do not always suffice, and various extensions or modifications may be called for depending on the phenomena under consideration.

In this paper, we provide one such example, where we investigate the phase separation behavior of polymer mixtures made of mesoscopic segments.
Our work has been motivated by the recent attempts to simulate chromatin organization in cellular nucleus. 
In ref.~\cite{Fujishiro_2022}, Fujishiro and Sasai constructed a polymer model of the whole genome of human cells, where each chromatin is modeled as a succession of soft-core monomers. 
Here, individual monomers (beads) represent $\sim 10^2$ kbp of DNA, which is much larger than the conventional monomers defined in standard theory or simulation of polymer systems. They argued that the interaction between such  mesoscopic segments is soft and repulsive, and the imbalance in such repulsion in systems with e.g., eu- and hetero-chromatic monomers would trigger the phase separation. 
Similar modelings of large scale behavior of chromatin with soft-core potential naturally arises after the coarse-graining, hence have been employed in other works as well~\cite{Das_2022, Das_2024, Kadam_2023, Michele_2016, Komoto_2022}, where the soft potential incorporates the entropic effect relevant to the mesoscipic segments.

How can we describe such a phase separation phenomena in chromatin theoretically? The immediate complication lies in the copolymer nature of the chromatin model~\cite{Erica_2020, Das_2022}. However, even if we let aside the sequence effect and restrict our attention to a binary mixture of homopolymers, the application of a Flory-Huggins theory is hampered because of the allowed overlap between monomers due to the soft-core nature of the inter-monomer potentials. 
A key insight would thus be obtained by the phase behavior of the mixture of soft particles. This problem has been extensively studied by groups of Likos, L\"owen and Kahl\cite{likos1998, likos2001, likos2007}.
Very recently, Sta\v{n}o, Likos and Egorov have extended the framework to the system of chains of soft beads\cite{stano2023}. Although their primal target is a mixture of linear polymers and ring polymers (or polycatenanes), we expect that the same physics applies to chromatin system, too.

Our first aim is thus to recapitulate and to numerically validate the theoretical framework for the binary mixture of polymers made of soft monomers, which allows one to analyze the phase behavior.
In Sec. ~\ref{Sec:f_blend}, we introduce the mean-field free energy for our system. From the analysis of the free energy, we present in Sec. ~\ref{PD_blend}, the miscibility phase diagram and compare it with the result from molecular dynamic (MD) simulations. Sec. ~\ref{Discussion} is devoted to discussions on universal aspects of the phase diagram, comparison with a conventional Flory-Huggins theory, and connection to the Gaussian core model. Building on the framework, we also discuss its extension to a system containing copolymers in Sec.~\ref{sec:RC}.

\section{Free energy of the soft repulsive polymer mixture}
\label{Sec:f_blend}
Following Sta\v{n}o ~\cite{stano2023}, we adopt the following free energy density for the mixture of polymers modeled as the succession of soft beads which represent mesoscopic segments
\begin{eqnarray}
\frac{f}{k_BT}& = &\frac{c_a}{N_a} \ln{c_a} +  \frac{c_b}{N_b} \ln{c_b}   \nonumber \\
&&+ \frac{1}{2} \chi_{aa} c_a^2 +  \frac{1}{2} \chi_{bb}  c_b^2+ \chi_{ab} c_a c_b
\label{f_soft}
\end{eqnarray}
where $k_BT$ is thermal energy, $c_x$ and $N_x$ are, respectively, the number density of beads and the chain length (number of beads per chain) of component $x (=a \ {\text or}\  b)$. Parameters $\chi_{xy} (>0)$ represent the strength of the repulsive interaction between beads $x$ and $y$. 

Note that in this representation, the parameters $\chi_{xy}$ have a unit of volume, and we measure them in unit of the volume of individual beads. In other words, we assume, for simplicity, the characteristic size ($\sigma$) of beads $a$ and $b$ are equal, which is taken to be the unit of length. Although there is no attraction, the phase separation may be induced by the asymmetry in the repulsion, i.e., $\chi_{aa} \neq \chi_{bb}$. At first sight, Eq.~(\ref{f_soft}) looks as a free energy in the second virial approximation valid for low concentrations. As we shall show below, however, the free energy~(\ref{f_soft}) is capable of describing the phase separation in concentrated regime $(c_a + c_b) \sigma^3 > 1$ (see Seq.~\ref{sec:GCM} for discussion). 

Let us first clarify a mathematical aspect relevant to the phase equilibria condition in the system described by the free energy~Eq. (\ref{f_soft}). If the homogeneous mixture of polymer A and polymer B separates into phase 1 and phase 2,  the demixed state is specified by the concentrations of both components in respective phases, i.e., $(c_a^{(1)},\  c_b^{(1)})$ and $(c_a^{(2)}, \ c_b^{(2)})$. The number of unknowns is thus $n_u=4$.  
On the other hand, the phase equilibria between two phases indicates the equalities of chemical potentials $\mu_x^{(\alpha)}=\partial f/\partial c_x^{(\alpha)}$ between two phases ($\alpha=1$ or $2$) for both components ($x=a$ or $b$), i.e., $\mu_a^{(1)}=\mu_a^{(2)}$ and $\mu_b^{(1)}=\mu_b^{(2)}$ and also the mechanical balance ensured by the equality of pressure  $P^{(1)}=P^{(2)}$ , where $P(c_a,c_b) = c_a [\partial f (c_a,c_b)/ \partial c_a] + c_b  [\partial f (c_a,c_b)/ \partial c_b]  - f(c_a,c_b)$, leading to the number of condition to determine the phase equilibria $n_c=3$.
Comparing the number of unknowns and that of conditions, we expect that the dimensionality of the phase boundary, i.e., binodal is $d_{pb}=n_u-n_c=1$.

\section{Phase diagram of the soft repulsive polymer mixture}
 \label{PD_blend}
\begin{figure}[h]
	\centering
	\includegraphics[width=0.45\textwidth]{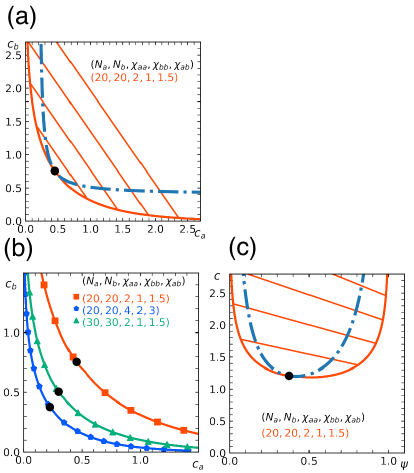}
	\caption{Miscibility phase diagram obtained from the free energy~(\ref{f_soft}). (a) Binodal (solid curve) and  spinodal (dashed curve) in the case of repulsion parameters $ \chi_{aa}=2$, $\chi_{bb}=1$, $\chi_{ab}=1.5$ and the chain length $N_a=N_b=20$. Two curves meet at the critical point marked by a circle. Some examples of tie lines are also shown. (b) Dependence of phase diagram on chain length and repulsion parameters, where binodals obtained from different set of parameters are shown.(c) Replot of the phase diagram (a) in the plane spanned by total concentration $c = c_a + c_b$ and the composition $\psi = c_a/c$.}
	\label{fig_PD_t}
			\vspace{0.2 cm}
\end{figure}

In Fig.~\ref{fig_PD_t} (a), we show an example of the miscibility phase diagram obtained from the free energy~Eq. (\ref{f_soft}) under the fixed interaction parameters.
As we have discussed, the phase diagram is two-dimensional spanned by $c_a$ and $c_b$, in which the uniform state (bottom left) and the demixed state (upper right) are separated by one-dimensional phase boundary. Tie lines, which connect $(c_a^{(1)},\  c_b^{(1)})$ and $(c_a^{(2)}, \ c_b^{(2)})$ in demixed state, are negatively sloped, indicating the phase separation is a segregative type. 
As expected, the region for demixing widens with the increase in either chain length or the repulsion strength, see Fig. ~\ref{fig_PD_t} (b).
Note that despite the symmetry in the chain length $N_a = N_b$ in the examples shown here, the phase diagram exhibits the asymmetry about the diagonal $c_a=c_b$. The asymmetry is caused by the difference in physical properties of type $a$ and $b$ beads, which leads to the phase rich in softer beads $b$ being more concentrated than the other. Such a feature can be made more evident by re-plotting the phase diagram in total concentration $c = c_a + c_b$ and the composition $\psi = c_a/c$ plane (Fig.~\ref{fig_PD_t} (c)).

To check the validity of the free energy prediction, we have performed numerical simulations of the polymer mixture.
Briefly, the system is a mixture of two types of linear homopolymers $A$ and $B$, where $A$ ($B$) polymer is made of a succession of $N_a$ ($N_b$) beads of type $a$ ($b$) (see Appendix for details of the simulation model). To represent the soft repulsion between monomers, we employ the Gaussian potential, see Eq.~(\ref{U_pair}) in Appendix, where the strength of the repulsion between x-bead and y-bead is $\epsilon_{xy}$ in unit of $k_BT$.

\begin{figure}[h]
	\centering
	\includegraphics[width=0.47\textwidth]{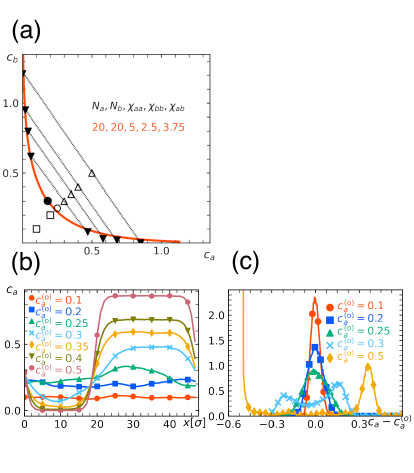}
	\caption{ Phase boundary (solid circle) of miscibility phase diagram obtained from numerical simulation with interaction parameters $ \epsilon_{aa}=2$,  $\epsilon_{bb} =1$,  $\epsilon_{ab}=1.5 $ and the chain length $N_a = N_b=20$. Open symbols (square, circle and triangle) indicate the overall concentrations of the system simulated; square or triangle indicates that the uniform state is stable or unstable, while circle represents the vicinity of the critical point. Overlapped solid curve is the binodal curve obtained from mean field theory with  $\chi_{aa}=5$,  $\chi_{bb}=2.5$, $\chi_{ab}=3.75$.  (b) Spatial profile of concentration of $a$-bead under various overall concentrations.(c) Histogram of number density of $a$-bead, where horizontal axis represents the deviation from the uniform cooncentration $c_a-c_a^{\rm{(o)}}$.}
	\label{fig_PD_s}
			\vspace{0.2 cm}
\end{figure}

The numerically determined phase boundary shown in Fig.~\ref{fig_PD_s} (a) is obtained with the interaction strength $ \epsilon_{aa}=2$,  $\epsilon_{bb} =1$,  $\epsilon_{ab}=1.5 $ and the chain length $N_a = N_b=20$, where the overall concentrations are varied from $(c_a^{\rm{(o)}}, c_b^{{\rm (o)}}) = (0.1, 0.1)$ to $(c_a^{\rm{(o)}}, c_b^{{\rm (o)}}) = (0.5, 0.5)$.
We find that the mixture is homogeneous at $(c_a^{\rm{(o)}}, c_b^{{\rm (o)}}) = (0.1, 0.1)$, but develops large concentration fluctuation at $(c_a^{\rm{(o)}}, c_b^{{\rm (o)}}) = (0.25, 0.25)$, and further increase in concentration leads to a well-defined phase separated structure (Fig.~\ref{fig_PD_s} (b)). 
Remarkably, the numerically determined phase diagram resembles that predicted by the free energy analysis (Fig.~\ref{fig_PD_t}). More specifically, we find that the numerical and analytical phase diagrams almost overlap under the correspondence $\chi_{xy}/ \epsilon_{xy} \simeq 2.5 $ between interaction parameters in free energy and interaction strengths in simulation.   

\section{Discussions}
\label{Discussion}
\subsection{General aspects of phase diagram}
\label{GeneralAspect}
In Sec.~\ref{PD_blend}, we have shown one example how the phase boundary alters  with the change in chain length or the interaction strength (Fig. ~\ref{fig_PD_t} (b)). To clarify the dependence of the shape of phase diagram on system parameters in a more systematic way, it is desirable to find out universal aspects inherent to the model described by the free energy~(\ref{f_soft}).

\begin{figure}[h]
	\centering
	\includegraphics[width=0.5\textwidth]{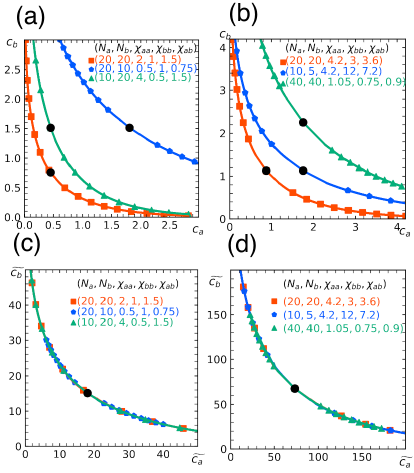}
	\caption{Rescaling of phase diagram.  Phase boundaries for three different conditions with common families specified by (a) $k_1=2$, $k_3=3/\sqrt{8}$ and (b) $k_1=1.4$, $k_3=6/\sqrt{35}$. In (a), parameters of one of the three conditions shown here are set to $(\chi_{aa}, \chi_{bb}, \chi_{ab}, N_a, N_b)=(2,1,1.5,20,20)$ and the other two are obtained from the transformations Eqs~(\ref{parameter_change1}) and~(\ref{parameter_change2}) with $p=0.25$, $q=1$, $k=4$ (blue) and $p=2$, $q=0.5$, $k=0.25$ (green). In (b), reference parameters are set to  $(\chi_{aa}, \chi_{bb}, \chi_{ab}, N_a, N_b)=(4.2,3,3.6,20,20)$ and the other two are obtained from the transformations Eqs~(\ref{parameter_change1}) and~(\ref{parameter_change2}) with $p=1$,$q=0.5$, $k=4$(blue) and $p=0.25$, $q=2$, $k=1$(green). (c) and (d) are master curves of the phase diagrams for the family with $k_1=2$, $k_3=3/\sqrt{8}$ and $k_1=1.4$, $k_3=6/\sqrt{35}$, respectively, upon rescaling of original phase diagrams (a) and (b).
	}
	\label{fig3}
			\vspace{0.2 cm}
\end{figure}

To this end, we introduce the rescaled concentrations ${\tilde c_a} = c_a N_a \chi_{aa}$ and ${\tilde c_b} = c_b N_b \chi_{bb}$, which enables us to rewrite Eq.~(\ref{f_soft}) as
\begin{eqnarray}
\frac{f}{k_BT} =  \frac{1}{N_a^2 \chi_{aa}} \left[{\tilde c_a} \ln{{\tilde c_a}} +  k_1 {\tilde c_b} \ln{{\tilde c_b}} + \frac{1}{2}  {\tilde c_a}^2 +  \frac{1}{2} k_1 {\tilde c_b}^2+ k_2 {\tilde c_a} {\tilde c_b} \right] \nonumber \\
\
\label{f_soft_rescaled}
\end{eqnarray}
where irrelevant linear terms in concentrations are dropped, and coefficients are
\begin{eqnarray}
k_1 = \frac{N_a^2}{N_b^2}\frac{\chi_{aa}}{\chi_{bb}}, \ k_2 = \frac{N_a}{N_b}\frac{\chi_{ab}}{\chi_{bb}} = \sqrt{k_1} \frac{\chi_{ab}}{\sqrt{\chi_{aa} \ \chi_{bb}}}
\label{parameter_comb}
\end{eqnarray}
The above free energy density is invariant under the parameter changes which keep $k_1$ and $k_3 = \chi_{ab}/\sqrt{\chi_{aa} \  \chi_{bb}}$ constant. These conditions are satisfied by the following transformations
\begin{eqnarray}
(\chi_{aa}, \chi_{bb}, \chi_{ab}) &\Rightarrow& p (\chi_{aa}, k \chi_{bb}, \sqrt{k} \chi_{ab}) \label{parameter_change1} \\
(N_a, N_b) &\Rightarrow& q (N_a, N_b/\sqrt{k}) \label{parameter_change2}
\label{parameter_change2}
\end{eqnarray}
where $p, q, k$ are positive real numbers.
Therefore, with the change in parameters according to Eqs~(\ref{parameter_change1}) and~(\ref{parameter_change2}), the phase diagram drawn in terms of rescaled concentrations remains the same.
A similar analysis has been done by Sta\v{n}o et. al~\cite{stano2023}.
Indeed, from the stability analysis of the uniform state, one finds the spinodal curve
\begin{eqnarray}
\frac{(1+{\tilde c_a})(1+{\tilde c_b})}{{\tilde c_a} {\tilde c_b}} = k_3^2
\label{eq:spinodal}
\end{eqnarray}
and the critical point is determined by Eq.~(\ref{eq:spinodal}) together with
\begin{eqnarray}
\frac{{\tilde c_a}(1+{\tilde c_a})^3}{ {\tilde c_b}(1+{\tilde c_b})^3} = k_1
\label{eq:cp}
\end{eqnarray}
Eq.~(\ref{eq:spinodal}) indicates that a necessary condition for the phase separation $k_3>1  \Leftrightarrow \chi_{ab} > \sqrt{\chi_{aa}\chi_{bb}}$, and Eq.~(\ref{eq:cp}) determines the location of critical point on the spinodal curve ~\cite{stano2023}.
In Fig.~\ref{fig3}, we demonstrate a collapse of the phase diagram upon rescaling. 

 \subsection{Comparison with Flory-Huggins theory}
 It is instructive to compare the present theory with the standard Flory-Huggins theory for polymer blends.
The Flory-Huggins free energy (per lattice site) for a blend of polymer A and B is written as~\cite{deGennes_scaling_concept}~\cite{doi1995introduction}
\begin{eqnarray}
\frac{f_{{\rm FH}}}{k_BT} = \frac{\phi_a}{N_a} \ln{\phi_a} +  \frac{\phi_b}{N_b} \ln{\phi_b}  + \chi \phi_a \phi_b
\label{FH_binary}
\end{eqnarray}
where $\phi_x$ is the volume fraction of component $x$, and a non-dimensional parameter $\chi$ measures the nature and the strength of interaction. The incompressiblity condition enforces $\phi_a + \phi_b =1$.
Since $\chi$ is usually positive, corresponding to attraction among the like-species, such an interaction acts as a driving force to the phase separation. The free energy~(\ref{FH_binary}) reduces to that of polymer solution in the limit $N_b=1$ where the component $b$ represents a solvent.

\begin{figure}[b]
	\centering
	\includegraphics[width=0.5\textwidth]{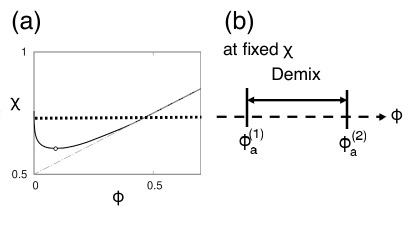}
	\caption{(a) Miscibility phase diagram for a polymer blend obtained from Flory-Huggins free energy Eq.~(\ref{FH_binary}). (b) Under a fixed $\chi$-parameter ($\chi > \chi_c$), the phase boundaries are points ($d_{{\rm pb}}=0$) on a line.}
	\label{fig4}
	\vspace{0.2 cm}
\end{figure}
 
Figure~\ref{fig4} shows  the phase diagram calculated from Eq.~(\ref{FH_binary}).
When we fix the interaction parameter at the value $\chi > \chi_c$, where $\chi_c=(\sqrt{1/Na} + \sqrt{1/Nb})^2/2$ is the critical value for the phase separation, the phase diagram as a function of $\phi_a$ is one-dimensional line with two points $\phi_a^{(1)}$ and $\phi_a^{(2)}$ representing the phase boundaries. If the overall concentration falls in between these two points, the uniform state is metastable (outside spinodal region) or unstable (inside spinodal region) and the system phase separates into the A-poor (dilute) and A-rich (concentrated) phases with the volume fractions $\phi_a^{(1)}$ and $\phi_a^{(2)}$, respectively. Note that the dimensionality of the phase boundary at fixed $\chi$ is $d_{{\rm pb}}=0$, i.e., points, which is a consequence of the equality of number of conditions ($n_c=2$, i.e.,  $\mu_a^{(1)}=\mu_a^{(2)}$ and $\mu_b^{(1)}=\mu_b^{(2)}$) and the number of unknowns ($n_u=2$, i.e., $\phi_a^{(1)}$ and $\phi_a^{(2)}$).
We note that a set of conditions $\mu_a^{(1)}=\mu_a^{(2)}$ and $\mu_b^{(1)}=\mu_b^{(2)}$ is equivalent to $\mu_a^{(1)}=\mu_a^{(2)}$ and $\Pi^{(1)}=\Pi^{(2)}$, where $\Pi(\phi_a) = \phi_a  [df_{{\rm FH}}(\phi_a)/d\phi_a] - f_{{\rm FH}}(\phi_a)$ is the osmotic pressure, the use of which may be more common in polymer solution, where the component $b$ is regarded solvent.
These two methods are equivalent due to the relation 
 \begin{eqnarray}
 -\Pi(\phi_a) v_0 = \mu_b(\phi_a) 
 \label{Pi_muB}
 \end{eqnarray}
 which follows from the incompressible condition~\cite{deGennes_scaling_concept}, where $v_0$ is the volume of the monomers and solvents.
 In contrast, in our description of polymer mixture with soft potential, the solvent degrees of freedom is already integrated out, and $c_a$ and $c_b$ are independent variables without constraint, i.e., free from the incompressible condition. One can conceive that our system under consideration is a three component system (two solute $A$ and $B$ plus solvent), and the free energy density~(\ref{f_soft}) represents a mesoscopic description after coarse-graining. 
 
 It is also known that the interaction part in the Flory-Huggins free energy initially takes the form $(\chi_{aa}/2) \phi_a^2 + (\chi_{bb}/2) \phi_b^2 + \chi_{ab} \phi_a \phi_b$. Rewriting it into the form of Eq.~(\ref{FH_binary}) with a single parameter $\chi = \chi_{aa}/2 + \chi_{bb}/2 - \chi_{ab}$ is made using the incompressibility condition. Again, it does not apply to our soft polymer description. Since our system originally possesses three components, we naturally need three interaction parameters to characterize the system.
 We also note that the critical $\chi$ parameter in blends of long polymers $N_A, N_B \gg 1$ is $\chi_c \rightarrow 0$ in Flory-Huggins theory. A necessary condition for the phase separation in this limit is thus $\chi > 0 \Leftrightarrow (\chi_{aa}+\chi_{bb})/2 > \chi_{ab}$. In contrast, as discussed in Sec.~\ref{Discussion}, the corresponding condition in our soft polymer mixtures is $\chi_{ab} > \sqrt{\chi_{aa} \chi_{bb}}$ independent of the chain length\cite{stano2023}.
 
\subsection{Relation with Gaussian core model}
\label{sec:GCM}
In our model, polymers are described as $N$ successive soft beads, where these beads are already mesoscopic entities with their internal degrees of freedom integrated out. We note here that, starting from a microscopic model, there is a freedom to choose $N$, i.e., the degree of the coarse-graining.
Although the extreme limit of the choice is $N=1$, in which individual polymers are described as single soft particles, the validity of such a description may break down at high concentration\cite{Pierleoni2007}.

It is known that the effective pair potential between two isolated polymer coils in dilute solution can be well approximated by a simple Gaussian potential
\begin{eqnarray}
\frac{U(r)}{k_BT} = \epsilon \exp{\left( -\frac{r^2}{R^2}\right)}
\label{U_G}
\end{eqnarray}
where $\epsilon \simeq 2$ and the width $R$ is of order of the gyration radius the coil ~\cite{Louis_2000}. The fact that the energy scale of the interaction is of order of thermal energy indicates entropic nature of the interaction. It has been shown that the above potential also provides a reasonable description for the effective interaction in semidilute solution, where polymers are overlapped.
Thermodynamic properties of fluid composed of soft particles interacting through Eq.~(\ref{U_G}), i.e., the Gaussian core model, has been analyzed in detail by Louis et. al.~\cite{Louis_2000}. 
Phase separation in binary mixture of such fluids has been also extensively studied~\cite{likos1998, likos2001, likos2007}.
As discussed in Sec.~\ref{GeneralAspect}, our free energy~(\ref{f_soft}) can formally be mapped to that case by $(N_a, N_b, c_a, c_b) \rightarrow (1,1,c_a/N_a, c_b/N_b)$. 
The analysis in Sec.~\ref{GeneralAspect} may then indicate that the miscibility phase diagram is intact if we simultaneously transform the interaction parameters as $(\chi_{aa}, \chi_{bb}, \chi_{ab}) \rightarrow (\chi_{aa}N_a^2, \chi_{bb}N_b^2, \chi_{ab}N_a N_b)$. 
One may then conclude that the introduction of the ``polymerization index" $N_a$ and $N_b$ might be auxiliary for the description of homopolymer mixtures. 
However, there are, at least, two reasons we need the polymeric description.
First, $N=1$ description is known to suffer from a significant concentration dependence of the effective repulsive interactions once polymer coils start to overlap deep in semidilute concentration regime. This motivates the multisegment description with $N>1$ ~\cite{Pierleoni2007}, where a suitable choice for $N$ would be guided by the overlapping condition for mesoscopic segments.
Second, once there arises a characteristic length scale in the problem, we need the polymeric description as a succession of beads with appropriate degree of coarse-graining. In Sec.~\ref{sec:RC}, we provide one such example, where we analyze the effect of modulation in local physical properties along polymers, (i.e., due to post translational modification) on the phase separation.

Another point deserving comment is the relation between the strength of the interaction potential $\epsilon_{xy}$ and the interaction parameter $\chi_{xy}$ in our free energy~(\ref{f_soft}). We have shown in Sec.\ref{PD_blend} that the simulation results quantitatively match with the free energy prediction under the relation $\chi_{xy}/ \epsilon_{xy} \simeq 2.5 $.
Since the free energy (\ref{f_soft}) takes apparently the same form as the virial expansion up to second order, one may expect that the $\epsilon_{xy} - \chi_{xy}$ relation would be obtained from $\chi_{xy}= - \int (e^{-U_{xy}(r)/k_BT}-1) d{\vec r}$.
As emphasized in ref.~\cite{Louis_2000}, however, the free energy~(\ref{f_soft}) is based on the random phase approximation (RPA)\cite{hansen2013theory}. As such, it becomes more and more accurate in higher concentration regimes in contrast to the second virial approximation\cite{hansen2013theory}. In fact, unlike the virial expansion, the quadratic form of the free energy in concentrations is a consequence of the RPA closure, where the direct correlation functions, which appears in the Ornstein-Zernike relation, is independent of the concentrations. The analysis of equation of state with RPA leads to the identification $\chi_{xy}= (1/k_BT) \int U_{xy}(r) d{\vec r} = \pi^{3/2} \sigma_{xy}^3 \epsilon_{xy}$.
Given the resultant ratio  $\chi_{xy}/ \epsilon_{xy} = \pi^{3/2} $ is considered to be an upper bound compared to a more accurate estimate e.g., obtained from hypernetted chain closure ~\cite{Louis_2000}, we find our result $\chi_{xy}/ \epsilon_{xy} \simeq 2.5 $ reasonable and providing an overall consistency of the soft core model description of the phase separation based on the free energy~Eq.(\ref{f_soft}).

\section{Mixtures with random copolymers}
\label{sec:RC}
So far discussed is a foundation for the coarse-grained description of polymer mixture, where polymers are represented as succession of soft mesoscopic beads. In particular, we have focused on the phase behavior of mixture of homopolymers.
As stated in Introduction,  however, one of the main motivations to necessitate such a description is its relevance to describe the large scale behavior of chromatin in cellular nucleus. In this section, we would like to discuss a simple extension of our theory, which may be linked to a certain aspect of chromatin organization in living cells.

It is known that interphase chromatin in early embryo is quite homogeneous inside nucleus, which is, in certain sense, reminiscent to a uniform solution of homopolymers~\cite{Sakaue_Kimura_PRL2022}. With the progress of the development stage, however, several characteristic structures, such as heterochromatin foci and transcriptional factories, start to appear~\cite{Arai_2017}. Responsible for such structure formations would be a phase separation, which is driven by local alternation of chromatin monomers caused by e.g., post-translational modification. The change in the chemical state in chromatin monomers likely induces the modulation of physical properties along chromatin polymer, which could be represented by a copolymer model. 
Since the variation in repulsive forces primarily reflects the difference in density of core-bearing monomers within the coarse-grained segments\cite{Fujishiro_2022, Maeshima_2024}, the segment ``a" represents regions where chromatin exists in a relaxed, less condensed state, reminiscent of euchromatin, while the segment ``b" corresponds to more condensed regions akin to heterochromatin. The structure formation under consideration could thus be treated by the appearance of copolymers in the matrix of homopolymers. With this in mind, let us consider a mixture of homopolymers H (with length $N_h$), which composes of type $a$ beads only, and copolymer $C$ (with length $N_c$) , which composes of types $a$ and $b$ beads.
The monomer concentration of $H$ and $C$ polymers are $c_h$ and $c_c$, respectively. For the analytical tractability in a simple mean-field description, we assume the latter to be a {\it random} copolymer, which is characterized by the fraction $\alpha$ of $b$ beads, i.e.,  the number of $b$ beads in a copolymer C is $\alpha N_c$.

The free energy of the mixture is written as
\begin{eqnarray}
\frac{f}{k_BT}& = &\frac{c_h}{N_h} \ln{c_h} +  \frac{c_c}{N_c} \ln{c_c}   \nonumber \\
&&+ \frac{1}{2} \chi_{hh} c_h^2 +  \frac{1}{2} \chi_{cc}  c_c^2+ \chi_{hc} c_h c_c
\label{f_soft_cop}
\end{eqnarray}
which takes the same form as Eq.~(\ref{f_soft}) except for the appearance of new interaction parameters. 
While $\chi_{hh}=\chi_{aa}$ trivially from the definition of the homopolymer H, the others $\chi_{cc}$ and $\chi_{hc}$ are nontrivial, which appear instead of $\chi_{bb}$ and $\chi_{ab}$ , respectively.
Given the randomness in the sequence of the copolymer, we can evaluate these interaction parameters as mean values of the inter-beads interactions  $\chi_{aa}$,  $\chi_{bb}$,  $\chi_{ab}$;
\begin{eqnarray}
\chi_{cc}&=&(1-\alpha)^2 \chi_{aa} + \alpha^2 \chi_{bb} + 2 \alpha (1-\alpha) \chi_{ab}  \label{chi_copolymer_1} \\
\chi_{hc}&=&(1-\alpha) \chi_{aa} + \alpha \chi_{ab}
\label{chi_copolymer_2}
\end{eqnarray}

\begin{figure}[H]
	\centering
	\includegraphics[width=0.5\textwidth]{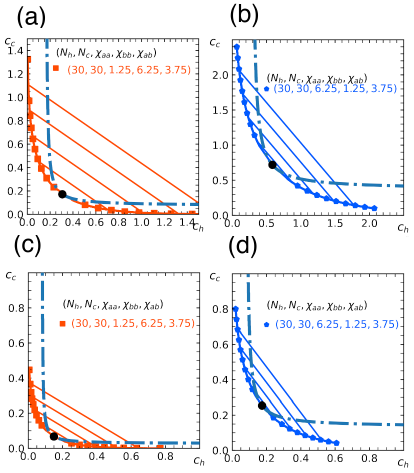}
	\caption{Miscibility phase diagram of homo- and copolymer mixtures obtained from the free energy~(\ref{f_soft_cop}). Binodal (solid curve) and  spinodal (dashed curve) for the chain length $N_h=N_c=30$. ts{Fraction of $b$ beads in copolymer is $\alpha=0.3$ in (a), (b) and $\alpha=0.5$ in (c), (d).  Repulsion parameters between beads are $ \chi_{aa}=1.25$, $\chi_{bb}=6.25$, $\chi_{ab}=3.75$ in (a), (c) and $ \chi_{aa}=6.25$, $\chi_{bb}=1.25$, $\chi_{ab}=3.75$ in (b), (d). Critical point (marked by a circle), and some examples of tie lines are also shown.} }
	\label{Fig5}
			\vspace{0.2 cm}
\end{figure}

\begin{figure}[H]
	\centering
\includegraphics[width=0.5\textwidth]{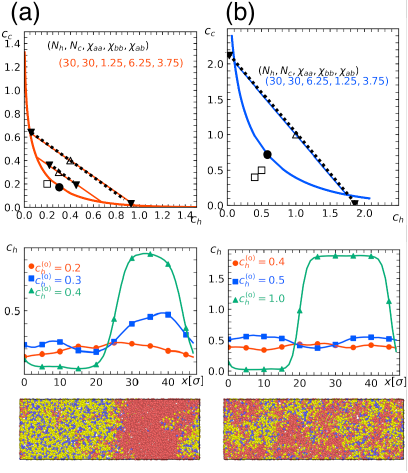}
	\caption{Quantitative comparison of theoretical phase diagram against numerical simulation. Chain lengths are $N_h=N_c=30$, the fraction of $b$ beads in copolymer is $\alpha=0.3$, and repulsion parameters between beads are (a) $ \chi_{aa}=1.25$, $\chi_{bb}=6.25$, $\chi_{ab}=3.75$ and (b) $ \chi_{aa}=6.25$, $\chi_{bb}=1.25$, $\chi_{ab}=3.75$. These conditions correspond to those in Fig.~\ref{Fig5} (a) and (b), respectively.  
 (Top) Binodal (solid curve) with the critical point marked by solid circle. Open symbols indicate the overall concentration $(c_h^{(0)},c_c^{(0)})$ adopted in numerical simulations performed under the parameter correspondence $\epsilon_{xy} = \chi_{xy}/2.5$, where open squares and triangles, respectively, indicate the one-phase and two-phase regions. In the latter, concentrations after phase separation are shown by solid triangles, which are connected by tie lines.
 (Middle) Spatial profiles of monomer concentration of H-polymer.
 (Bottom) Typical snapshots obtained in simulations with $(c_h^{(0)},c_c^{(0)})=(0.4, 0.4)$, where red beads represent type-a beads contained in H-polymer, while yellow or blue beads represent type-a or type-b beads contained in C-polymer, respectively. The snapshots were rendered using OVITO\cite{OVITO}. }
	\label{Fig6}
			\vspace{0.2 cm}
\end{figure}

\begin{figure}[H]
	\centering
\includegraphics[width=0.5\textwidth]{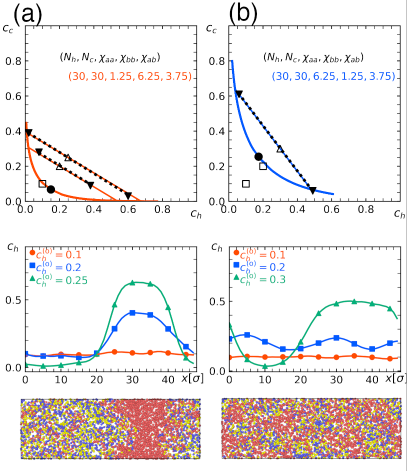}
	\caption{Quantitative comparison of theoretical phase diagrams against numerical simulation. Parameters are the same as those in Fig.~\ref{Fig6} except for the fraction of $b$ beads in copolymer, which is $\alpha=0.5$ here. These conditions correspond to those in Fig.~\ref{Fig5} (c) and (d), recpectively.  
 (Top) Binodal (solid curve) with the critical point marked by solid circle. Open symbols indicate the overall concentration $(c_h^{(0)},c_c^{(0)})$ adopted in numerical simulations performed under the parameter correspondence $\epsilon_{xy} = \chi_{xy}/2.5$, where open squares and triangles, respectively, indicate the one-phase and two-phase regions. In the latter, concentrations after phase separation are shown by solid triangles, which are connected by tie lines.
 (Middle) Spatial profile of monomer concentration of H-polymer.
 (Bottom) Typical snapshots obtained in simulations with $(c_h^{(0)},c_c^{(0)})=(0.2, 0.2)$, where red beads represent type-a beads contained in H-polymer, while yellow or blue beads represent type-a or type-b beads contained in C-polymer, respectively. The snapshots were rendered using OVITO\cite{OVITO}. }
	\label{Fig7}
			\vspace{0.2 cm}
\end{figure}

In Fig.~\ref{Fig5}, we show phase diagrams obtained from the free energy~(\ref{f_soft_cop}) with Eqs.~(\ref{chi_copolymer_1}) and~(\ref{chi_copolymer_2}) for a fixed $\alpha$. Note that $\alpha=0$ reduces to a homopolymer solution (with only type a bead), and $\alpha=1$ corresponds to a blend of homopolymers A and B analyzed in earlier sections. Here we show the cases for $\alpha=0.3,\ 0.5$. As expected, the region for phase separation enlarges with the fraction $\alpha$. In addition, the results depend on a relative stiffness between beads a and b.
As shown, the system is more prone to phase separation when the matrix polymer is softer $\chi_{aa} < \chi_{bb}$ reflecting the asymmetry in the phase diagram of homopolymer mixtures (Sec.~\ref{PD_blend}).

To check the validity of the free energy prediction, we again performed numerical simulations using Gaussian potentials to represent bead-bead soft repulsions. In Figs.~\ref{Fig6} and~\ref{Fig7}, we compare the theoretical phase diagram in Fig.~\ref{Fig5} with simulation results, where the repulsion strengths for Gaussian potentials are set to be $\epsilon_{xy} = \chi_{xy}/2.5$ between beads $x$ and $y$ as determined from the result of homopolymer mixtures. As shown, the agreement is rather satisfactory, demonstrating that the overall trend of phase separation is well captured by the proposed free energy. In Figs.~\ref{Fig6} and~\ref{Fig7}, we also show the spatial profiles of the monomer concentration $c_h$ of homopolymer together with the corresponding typical snapshots.

\section{Discussions and Summary}
Numerical simulations of large scale chromatin organization in cellular nucleus often adopt highly coarse-grained models, in which chromatin polymer is represented as a succession of soft beads~\cite{Fujishiro_2022, Das_2022,Das_2024,Kadam_2023,Michele_2016,Komoto_2022}. Unlike a model which employs nucleosomes as monomers of chromatin polymer, each of beads here represents a substantial amount of nucleosomes, thus regarded as a mesoscopic entity, allowing mutual overlaps with entropic penalty. It has been shown that the effective interaction between such mesoscipic segments is soft and repulsive, qualitative feature of which is well approximated by the Gaussian potential~\cite{Fujishiro_2022}.

We have considered binary mixtures of such soft repulsive polymers and investigated how the imbalance in repulsive interactions between different species leads to the phase separation. 
After summarizing universal aspects of the phase diagram based on invariant property of the free energy upon changes in parameter values, we have  extended the theory to mixtures including random copolymers, which may have some implications to chromatin phase separation. 

As discussed in Sec.~\ref{sec:RC}, random copolymer model is inspired by epigenetic modification of chromatin. This modification is performed and maintained by enzymes, thus includes energy consuming nonequilibrium process. In this sense, our description based on equilibrium framework should be considered a useful effective description to elucidate the impact of phase separation on chromatin organization. The same remark applies to many of current biophysical modelings of chromatin, not only for its structural organization but also for its dynamics. Yet, there are several other works, which emphasize possible impacts of various nonequilibrium effects on chromatin\cite{Zidovska2020, Put2019, BRUINSMA2014, Vandebroek2015, Sakaue_2016,Michieletto_2019, Cristian_2024, Iurii_2022, Akinori_2014}. 
Perhaps, some of these effects associated with nonequilibrium activities could be described by effective equilibrium models. Such a strategy may well work to understand some aspect of the problem, but may fail to capture other aspects.  In our opinion, it remains to be seen how and when nonequilibrium factors are critically important in chromatin biophysics.
The same comment would apply to topological constraints, another  factor presumably important in chromatin, but not explicitly included in our description~\cite{Halverson_2014, Angelo_2008}.  
In this regard, it is interesting to note that as discussed in~\cite{Sakaue_2016},  the similar physics as described in present paper may be important in blend of non-concatenated ring polymers, where the soft repulsion arises from the so-called topological volume due to topological constraints~\cite{Sakaue_2011, Frank-Kamenetskii1975, Sakaue_2018}.

As possible extensions of our work, we first note that our theory deals with the macro-phase separation, hence does not capture the possible appearance of mesophases. However, the occurrence of  the ``micro-phase separation" is naturally expected in copolymer systems, the elucidation of which should provide further insight into the problem of chromatin organization in nuclei.
Secondly, although we have only analyzed bulk property based on mean-field theory,  we expect that the effect of correlations and interfacial properties at phase boundary and near a confining wall could be analyzed by following an approach outlined in ref.~\cite{stano2023}. It would be interesting to see how such an analysis can be compared to the chromatin spatial profile, e.g. near the nuclear membrane.

Finally, we point out that there are several studies on compressibility effects in polymer solutions, which become evident, for instance, in pressure-induced phase separation ~\cite{SHAHAMAT2015100}. Here, interesting phenomena such as the acousto-spinodal decomposition have been predicted ~\cite{Rausouli2011}. Although comparison with these studies may be interesting, we note that the loss of incompressible condition in our description results after coarse-graining, i.e., integrating out solvent degrees of freedom. Therefore, to address the kinetic effect, we need to take properly solvent effects into account.

\section*{Appendix}
\subsection*{Simulation model}
The system is a mixture of two types of linear homopolymers $A$ and $B$, where $A$ ($B$) polymer is made of a succession of $N_a$ ($N_b$) beads of type $a$ ($b$). 
The potential energy in the system has two contributions. The first is the intrachain bonding potential
\begin{eqnarray}
\frac{U^{{\rm (b)}}(r)}{k_BT} = \frac{1}{2} k_b (r-r_0)^2
\label{U_bond}
\end{eqnarray}
which acts the bonding pairs to maintain the linear connectivity of the chain, where $r$ and $r_0$ denotes the separation between bead centers and the natural bond length, respectively. We set the spring constant $k_b = 70.0/ \sigma^2$ to keep the bond length nearly constant $ r_0 =  \ \sigma$, where $\sigma$ is the unit of the length (see below). The thermal energy, $k_BT$, is chosen as the unit energy in the simulation system. The second is the non-bonded interaction potential, which represents the soft repulsion between monomers. We employ the Gaussian core potential; the pair potential between one bead with type $X$ and another bead with type $Y$ reads
\begin{eqnarray}
\frac{U^{({\rm int)}}_{xy}(r)}{k_BT} =  \epsilon_{xy} \exp{\left(-\frac{r^2}{\sigma_{xy}^2} \right)}
\label{U_pair}
\end{eqnarray}
where $\epsilon_{xy}$ and $\sigma_{xy}$, respectively, measure the strength and the range of repulsive interaction between $x$ and $y$ beads. For simplicity, we set the range of repulsive interaction for all types of pair equal, i.e., $\sigma_{aa} = \sigma_{bb} = \sigma_{ab} = \sigma$, and adopt this range (denoted as $\sigma$) as the unit length. 
Note that this interaction is acting on all the bead pairs except for the nearest neighbors along the chain (bonded pairs).

 Molecular Dynamics (MD) simulations with fixed volume and the constant temperature are performed using the LAMMPS package~\cite{LAMMPS}. To integrate the equations of motion, we adopt the velocity Verlet algorithm, in which all beads are coupled to a Langevin thermostat with the damping constant $\gamma = \tau_0^{-1}$ with $\tau_0 = \sigma(m/k_BT)^{1/2}$, where 
 $m$ is the bead mass (assumed to be the same for type $a$ and $b$). This $\tau_0$ is chosen as the unit time. The integration time step is set at $0.01 \tau_0$. $( L_x, L_y, L_z )$ represents the size of the rectangular parallelepiped system box with periodic boundary conditions. This system box is placed in $- L_\omega / 2 < \omega < L_\omega / 2$, where $\omega$ represents the Cartesian axes $x, y,$ and $z$. $( L_x = 48 \sigma, L_y = 16 \sigma, L_z = 16 \sigma)$ is fixed unless $( L_x = 120 \sigma, L_y = 16 \sigma, L_z = 16 \sigma)$ is utlized for the simulation of the $A$-$B$ homopolymer mixtures at $( c_a, c_b ) = ( 0.3, 0.3 )$.

To prepare the initial state, we start from dilute solution, where the same numbers of the homopolymers $A$ and $B$ are distributed in a large cubic box with the size $( 200 \sigma, 200 \sigma, 200 \sigma )$. We run the simulation at $\epsilon_{aa} = \epsilon_{bb} = \epsilon_{ab} =$ $2.0$ (mixture of homopolymers) or $2.5$ (mixture with random copolymers) for $2 \times 10^7$ steps with slowly compressing the system box into the final system size $( L_x, L_y, L_z )$. In this way, we obtain the desired concentration of polymer mixture, where both the $A$ and $B$- polymers are homogeneously mixed. For the mixture with copolymers, $(1-\alpha)  N_c$ beads in each $B$-homopolymer in this initial configuration are randomly chosen, and turned into type-$a$ beads. We then set the interaction strengths to appropriate values in the sebsequent production run (see below). 
Then, we reassign the monomer label to adjust the initial spatial concentration profile to prepare the phase separated initial state.


To perform various statistical analysis, we sampled microstates of the system every 1000 steps after the system reaches equilibrium, i.e., at $2 \times 10^6$ steps (simulation runs in Sec. 4) and $3 \times 10^6$ steps (Sec. 5) after setting the interaction strengths to appropriate values.
However, for the case of simulation runs given in Sec. 4 at $( c_a^{(0)}, c_b^{(0)} ) = ( 0.3, 0.3 )$, the sampling starts at $9 \times 10^6$ steps, and for the case of Sec. 5 at $\alpha = 0.3, (\epsilon_{aa}, \epsilon_{bb}, \epsilon_{ab}) = (2.5, 0.5, 1.5), (c_h^{(0)}, c_c^{(0)}) = ( 0.5, 0.5 )$, the sampling starts at $7 \times 10^6$ steps.
In all simulations, we collect 1001 independent samples of particle configurations in equilibrium. For the simulation at $( c_a^{(0)}, c_b^{(0)} ) = ( 0.3, 0.3 )$, as only one exception, production runs end at $18 \times 10^6$ steps, and particle coordinates are sampled every 1000 steps from $9 \times 10^6$ to $18 \times 10^6$ steps, during which 9001 independent samples of particle configurations in equilibrium are collected for statistical accuracy improvement in the vicinity of the critical point. We have confirmed that the physical properties of the simulation system are not significantly changed when we start the simulation from the homogeneously mixed initial states.

\section*{Conflicts of interest}
`There are no conflicts to declare.

\section*{Acknowledgements}
We thank M. Sasai and S. Fujishiro for discussions. This work is supported by JSPS KAKENHI (Grants No. JP23H00369, JP23H04290 and JP24K00602).



\balance


\bibliography{phaseseparation} 

\providecommand*{\mcitethebibliography}{\thebibliography}
\csname @ifundefined\endcsname{endmcitethebibliography}
{\let\endmcitethebibliography\endthebibliography}{}
\begin{mcitethebibliography}{53}
\providecommand*{\natexlab}[1]{#1}
\providecommand*{\mciteSetBstSublistMode}[1]{}
\providecommand*{\mciteSetBstMaxWidthForm}[2]{}
\providecommand*{\mciteBstWouldAddEndPuncttrue}
  {\def\EndOfBibitem{\unskip.}}
\providecommand*{\mciteBstWouldAddEndPunctfalse}
  {\let\EndOfBibitem\relax}
\providecommand*{\mciteSetBstMidEndSepPunct}[3]{}
\providecommand*{\mciteSetBstSublistLabelBeginEnd}[3]{}
\providecommand*{\EndOfBibitem}{}
\mciteSetBstSublistMode{f}
\mciteSetBstMaxWidthForm{subitem}
{(\emph{\alph{mcitesubitemcount}})}
\mciteSetBstSublistLabelBeginEnd{\mcitemaxwidthsubitemform\space}
{\relax}{\relax}

\bibitem[de~Gennes(1979)]{deGennes_scaling_concept}
P.~G. de~Gennes, \emph{Scaling Concepts in Polymer Physics}, Cornell University Press, Ithaca, NY, 1979\relax
\mciteBstWouldAddEndPuncttrue
\mciteSetBstMidEndSepPunct{\mcitedefaultmidpunct}
{\mcitedefaultendpunct}{\mcitedefaultseppunct}\relax
\EndOfBibitem
\bibitem[Doi(1995)]{doi1995introduction}
M.~Doi, \emph{Introduction to Polymer Physics}, Oxford University Press, 1995\relax
\mciteBstWouldAddEndPuncttrue
\mciteSetBstMidEndSepPunct{\mcitedefaultmidpunct}
{\mcitedefaultendpunct}{\mcitedefaultseppunct}\relax
\EndOfBibitem
\bibitem[BATES(1991)]{BATES_1991}
F.~BATES, \emph{SCIENCE}, 1991, \textbf{251}, 898--905\relax
\mciteBstWouldAddEndPuncttrue
\mciteSetBstMidEndSepPunct{\mcitedefaultmidpunct}
{\mcitedefaultendpunct}{\mcitedefaultseppunct}\relax
\EndOfBibitem
\bibitem[{van de Witte} \emph{et~al.}(1996){van de Witte}, Dijkstra, {van den Berg}, and Feijen]{Witte_1996}
P.~{van de Witte}, P.~Dijkstra, J.~{van den Berg} and J.~Feijen, \emph{Journal of Membrane Science}, 1996, \textbf{117}, 1--31\relax
\mciteBstWouldAddEndPuncttrue
\mciteSetBstMidEndSepPunct{\mcitedefaultmidpunct}
{\mcitedefaultendpunct}{\mcitedefaultseppunct}\relax
\EndOfBibitem
\bibitem[Yin \emph{et~al.}(2022)Yin, Yang, and Wang]{Yin_2022}
X.~Yin, J.~Yang and H.~Wang, \emph{Advanced Functional Materials}, 2022, \textbf{32}, 2202071\relax
\mciteBstWouldAddEndPuncttrue
\mciteSetBstMidEndSepPunct{\mcitedefaultmidpunct}
{\mcitedefaultendpunct}{\mcitedefaultseppunct}\relax
\EndOfBibitem
\bibitem[Goh and Lane(2022)]{Choon_2022}
C.~F. Goh and M.~E. Lane, \emph{Advanced Drug Delivery Reviews}, 2022, \textbf{180}, 114077\relax
\mciteBstWouldAddEndPuncttrue
\mciteSetBstMidEndSepPunct{\mcitedefaultmidpunct}
{\mcitedefaultendpunct}{\mcitedefaultseppunct}\relax
\EndOfBibitem
\bibitem[Sakuta \emph{et~al.}(2020)Sakuta, Fujita, Hamada, Hayashi, Takiguchi, Tsumoto, and Yoshikawa]{Sakuta_2020}
H.~Sakuta, F.~Fujita, T.~Hamada, M.~Hayashi, K.~Takiguchi, K.~Tsumoto and K.~Yoshikawa, \emph{ChemBioChem}, 2020, \textbf{21}, 3323--3328\relax
\mciteBstWouldAddEndPuncttrue
\mciteSetBstMidEndSepPunct{\mcitedefaultmidpunct}
{\mcitedefaultendpunct}{\mcitedefaultseppunct}\relax
\EndOfBibitem
\bibitem[Brangwynne \emph{et~al.}(2009)Brangwynne, Eckmann, Courson, Rybarska, Hoege, Gharakhani, Juelicher, and Hyman]{Brangwynne_2009}
C.~P. Brangwynne, C.~R. Eckmann, D.~S. Courson, A.~Rybarska, C.~Hoege, J.~Gharakhani, F.~Juelicher and A.~A. Hyman, \emph{SCIENCE}, 2009, \textbf{324}, 1729--1732\relax
\mciteBstWouldAddEndPuncttrue
\mciteSetBstMidEndSepPunct{\mcitedefaultmidpunct}
{\mcitedefaultendpunct}{\mcitedefaultseppunct}\relax
\EndOfBibitem
\bibitem[Chen and Belmont(2019)]{Chen_2019}
Y.~Chen and A.~S. Belmont, \emph{CURRENT OPINION IN GENETICS \& DEVELOPMENT}, 2019, \textbf{55}, 91--99\relax
\mciteBstWouldAddEndPuncttrue
\mciteSetBstMidEndSepPunct{\mcitedefaultmidpunct}
{\mcitedefaultendpunct}{\mcitedefaultseppunct}\relax
\EndOfBibitem
\bibitem[Esposito \emph{et~al.}(2022)Esposito, Abraham, Conte, Vercellone, Prisco, Bianco, and Chiariello]{Esposito_2022}
A.~Esposito, A.~Abraham, M.~Conte, F.~Vercellone, A.~Prisco, S.~Bianco and A.~M. Chiariello, \emph{Polymers}, 2022, \textbf{14}, 1918\relax
\mciteBstWouldAddEndPuncttrue
\mciteSetBstMidEndSepPunct{\mcitedefaultmidpunct}
{\mcitedefaultendpunct}{\mcitedefaultseppunct}\relax
\EndOfBibitem
\bibitem[Yanagisawa(2022)]{Yanagisawa_2022}
M.~Yanagisawa, \emph{BIOPHYSICAL REVIEWS}, 2022, \textbf{14}, 1093--1103\relax
\mciteBstWouldAddEndPuncttrue
\mciteSetBstMidEndSepPunct{\mcitedefaultmidpunct}
{\mcitedefaultendpunct}{\mcitedefaultseppunct}\relax
\EndOfBibitem
\bibitem[Falk \emph{et~al.}(2019)Falk, Feodorova, Naumova, Imakaev, Lajoie, Leonhardt, Joffe, Dekker, Fudenberg, Solovei, and Mirny]{Falk_2019}
M.~Falk, Y.~Feodorova, N.~Naumova, M.~Imakaev, B.~R. Lajoie, H.~Leonhardt, B.~Joffe, J.~Dekker, G.~Fudenberg, I.~Solovei and L.~A. Mirny, \emph{Nature}, 2019, \textbf{570}, 395 -- 399\relax
\mciteBstWouldAddEndPuncttrue
\mciteSetBstMidEndSepPunct{\mcitedefaultmidpunct}
{\mcitedefaultendpunct}{\mcitedefaultseppunct}\relax
\EndOfBibitem
\bibitem[Riback \emph{et~al.}(2020)Riback, Zhu, Ferrolino, Tolbert, Mitrea, Sandersr, Wei, Kriwacki, and Brangwynne]{Riback_2020}
J.~A. Riback, L.~Zhu, M.~C. Ferrolino, M.~Tolbert, D.~M. Mitrea, D.~W. Sandersr, M.-T. Wei, R.~W. Kriwacki and C.~P. Brangwynne, \emph{Nature}, 2020, \textbf{581}, 209–214\relax
\mciteBstWouldAddEndPuncttrue
\mciteSetBstMidEndSepPunct{\mcitedefaultmidpunct}
{\mcitedefaultendpunct}{\mcitedefaultseppunct}\relax
\EndOfBibitem
\bibitem[Deviri and Safran(2021)]{Deviri_2021}
D.~Deviri and S.~A. Safran, \emph{Proc.Natl.Acad.Sci.USA}, 2021, \textbf{118}, e2100099118\relax
\mciteBstWouldAddEndPuncttrue
\mciteSetBstMidEndSepPunct{\mcitedefaultmidpunct}
{\mcitedefaultendpunct}{\mcitedefaultseppunct}\relax
\EndOfBibitem
\bibitem[Yamazaki \emph{et~al.}(2022)Yamazaki, Takagi, Kosako, Hirano, and Yoshimura]{Yamazaki_2022}
H.~Yamazaki, M.~Takagi, H.~Kosako, T.~Hirano and S.~H. Yoshimura, \emph{Nat. Cell Biol.}, 2022, \textbf{24}, 625–632\relax
\mciteBstWouldAddEndPuncttrue
\mciteSetBstMidEndSepPunct{\mcitedefaultmidpunct}
{\mcitedefaultendpunct}{\mcitedefaultseppunct}\relax
\EndOfBibitem
\bibitem[Yamamoto \emph{et~al.}(2023)Yamamoto, Yamazaki, Ninomiya, and Hirose]{Yamamoto_2023}
T.~Yamamoto, T.~Yamazaki, K.~Ninomiya and T.~Hirose, \emph{Commun. Biol.}, 2023, \textbf{6}, 1129\relax
\mciteBstWouldAddEndPuncttrue
\mciteSetBstMidEndSepPunct{\mcitedefaultmidpunct}
{\mcitedefaultendpunct}{\mcitedefaultseppunct}\relax
\EndOfBibitem
\bibitem[Hilbert \emph{et~al.}(2021)Hilbert, Sato, Kuznetsova, Bianucci, Kimura, Jülicher, Honigmann, Zaburdaev, and Vastenhouw]{Lennart_2021}
L.~Hilbert, Y.~Sato, K.~Kuznetsova, T.~Bianucci, H.~Kimura, F.~Jülicher, A.~Honigmann, V.~Zaburdaev and N.~L. Vastenhouw, \emph{Nat. Commun.}, 2021, \textbf{12}, 1360\relax
\mciteBstWouldAddEndPuncttrue
\mciteSetBstMidEndSepPunct{\mcitedefaultmidpunct}
{\mcitedefaultendpunct}{\mcitedefaultseppunct}\relax
\EndOfBibitem
\bibitem[Michieletto \emph{et~al.}(2019)Michieletto, Col\`{\i}, Marenduzzo, and Orlandini]{Michieletto_2019}
D.~Michieletto, D.~Col\`{\i}, D.~Marenduzzo and E.~Orlandini, \emph{Phys. Rev. Lett.}, 2019, \textbf{123}, 228101\relax
\mciteBstWouldAddEndPuncttrue
\mciteSetBstMidEndSepPunct{\mcitedefaultmidpunct}
{\mcitedefaultendpunct}{\mcitedefaultseppunct}\relax
\EndOfBibitem
\bibitem[Koyama \emph{et~al.}()Koyama, Iso, Norizoe, Sakaue, and Yoshimura]{Koyama_2024}
T.~Koyama, N.~Iso, Y.~Norizoe, T.~Sakaue and S.~H. Yoshimura, \emph{Charge block-driven liquid-liquid phase separation: mechanism and biological roles}, submitted\relax
\mciteBstWouldAddEndPuncttrue
\mciteSetBstMidEndSepPunct{\mcitedefaultmidpunct}
{\mcitedefaultendpunct}{\mcitedefaultseppunct}\relax
\EndOfBibitem
\bibitem[Fujishiro and Sasai(2022)]{Fujishiro_2022}
S.~Fujishiro and M.~Sasai, \emph{Proc. Natl. Acad. Sci. USA}, 2022, \textbf{119}, e2109838119\relax
\mciteBstWouldAddEndPuncttrue
\mciteSetBstMidEndSepPunct{\mcitedefaultmidpunct}
{\mcitedefaultendpunct}{\mcitedefaultseppunct}\relax
\EndOfBibitem
\bibitem[Das \emph{et~al.}(2022)Das, Sakaue, Shivashankar, Prost, and Hiraiwa]{Das_2022}
R.~Das, T.~Sakaue, G.~Shivashankar, J.~Prost and T.~Hiraiwa, \emph{eLife}, 2022, \textbf{11}, e79901\relax
\mciteBstWouldAddEndPuncttrue
\mciteSetBstMidEndSepPunct{\mcitedefaultmidpunct}
{\mcitedefaultendpunct}{\mcitedefaultseppunct}\relax
\EndOfBibitem
\bibitem[Das \emph{et~al.}(2024)Das, Sakaue, Shivashankar, Prost, and Hiraiwa]{Das_2024}
R.~Das, T.~Sakaue, G.~V. Shivashankar, J.~Prost and T.~Hiraiwa, \emph{Phys. Rev. Lett.}, 2024, \textbf{132}, 058401\relax
\mciteBstWouldAddEndPuncttrue
\mciteSetBstMidEndSepPunct{\mcitedefaultmidpunct}
{\mcitedefaultendpunct}{\mcitedefaultseppunct}\relax
\EndOfBibitem
\bibitem[Kadam \emph{et~al.}(2023)Kadam, Kumari, Manivannan, Dutta, Mitra, and Padinhateeri]{Kadam_2023}
S.~Kadam, K.~Kumari, V.~Manivannan, S.~Dutta, M.~K.~K. Mitra and R.~Padinhateeri, \emph{NATURE COMMUNICATIONS}, 2023, \textbf{14}, 4108\relax
\mciteBstWouldAddEndPuncttrue
\mciteSetBstMidEndSepPunct{\mcitedefaultmidpunct}
{\mcitedefaultendpunct}{\mcitedefaultseppunct}\relax
\EndOfBibitem
\bibitem[Pierro \emph{et~al.}(2016)Pierro, Zhang, Aiden, Wolynes, and Onuchic]{Michele_2016}
M.~D. Pierro, B.~Zhang, E.~L. Aiden, P.~G. Wolynes and J.~N. Onuchic, \emph{Proceedings of the National Academy of Sciences}, 2016, \textbf{113}, 12168--12173\relax
\mciteBstWouldAddEndPuncttrue
\mciteSetBstMidEndSepPunct{\mcitedefaultmidpunct}
{\mcitedefaultendpunct}{\mcitedefaultseppunct}\relax
\EndOfBibitem
\bibitem[Komoto \emph{et~al.}(2022)Komoto, Fujii, and Awazu]{Komoto_2022}
T.~Komoto, M.~Fujii and A.~Awazu, \emph{Biophysics and Physicobiology}, 2022, \textbf{19}, e190018\relax
\mciteBstWouldAddEndPuncttrue
\mciteSetBstMidEndSepPunct{\mcitedefaultmidpunct}
{\mcitedefaultendpunct}{\mcitedefaultseppunct}\relax
\EndOfBibitem
\bibitem[Hildebrand and Dekker(2020)]{Erica_2020}
E.~M. Hildebrand and J.~Dekker, \emph{Trends in Biochemical Sciences}, 2020, \textbf{45}, 385--396\relax
\mciteBstWouldAddEndPuncttrue
\mciteSetBstMidEndSepPunct{\mcitedefaultmidpunct}
{\mcitedefaultendpunct}{\mcitedefaultseppunct}\relax
\EndOfBibitem
\bibitem[Likos \emph{et~al.}(1998)Likos, L\"owen, Watzlawek, Abbas, Jucknischke, Allgaier, and Richter]{likos1998}
C.~N. Likos, H.~L\"owen, M.~Watzlawek, B.~Abbas, O.~Jucknischke, J.~Allgaier and D.~Richter, \emph{Phys. Rev. Lett.}, 1998, \textbf{80}, 4450--4453\relax
\mciteBstWouldAddEndPuncttrue
\mciteSetBstMidEndSepPunct{\mcitedefaultmidpunct}
{\mcitedefaultendpunct}{\mcitedefaultseppunct}\relax
\EndOfBibitem
\bibitem[Likos(2001)]{likos2001}
C.~N. Likos, \emph{Physics Reports}, 2001, \textbf{348}, 267--439\relax
\mciteBstWouldAddEndPuncttrue
\mciteSetBstMidEndSepPunct{\mcitedefaultmidpunct}
{\mcitedefaultendpunct}{\mcitedefaultseppunct}\relax
\EndOfBibitem
\bibitem[Likos \emph{et~al.}(2007)Likos, Mladek, Gottwald, and Kahl]{likos2007}
C.~N. Likos, B.~M. Mladek, D.~Gottwald and G.~Kahl, \emph{The Journal of Chemical Physics}, 2007, \textbf{126}, 224502\relax
\mciteBstWouldAddEndPuncttrue
\mciteSetBstMidEndSepPunct{\mcitedefaultmidpunct}
{\mcitedefaultendpunct}{\mcitedefaultseppunct}\relax
\EndOfBibitem
\bibitem[Staňo \emph{et~al.}(2023)Staňo, Likos, and Egorov]{stano2023}
R.~Staňo, C.~N. Likos and S.~A. Egorov, \emph{Macromolecules}, 2023, \textbf{56}, 8168--8182\relax
\mciteBstWouldAddEndPuncttrue
\mciteSetBstMidEndSepPunct{\mcitedefaultmidpunct}
{\mcitedefaultendpunct}{\mcitedefaultseppunct}\relax
\EndOfBibitem
\bibitem[Pierleoni \emph{et~al.}(2007)Pierleoni, Capone, and Hansen]{Pierleoni2007}
C.~Pierleoni, B.~Capone and J.-P. Hansen, \emph{The Journal of Chemical Physics}, 2007, \textbf{127}, 171102\relax
\mciteBstWouldAddEndPuncttrue
\mciteSetBstMidEndSepPunct{\mcitedefaultmidpunct}
{\mcitedefaultendpunct}{\mcitedefaultseppunct}\relax
\EndOfBibitem
\bibitem[Louis \emph{et~al.}(2000)Louis, Bolhuis, and Hansen]{Louis_2000}
A.~A. Louis, P.~G. Bolhuis and J.~P. Hansen, \emph{Phys. Rev. E}, 2000, \textbf{62}, 7961--7972\relax
\mciteBstWouldAddEndPuncttrue
\mciteSetBstMidEndSepPunct{\mcitedefaultmidpunct}
{\mcitedefaultendpunct}{\mcitedefaultseppunct}\relax
\EndOfBibitem
\bibitem[Hansen and McDonald(2013)]{hansen2013theory}
J.~Hansen and I.~McDonald, \emph{Theory of Simple Liquids: with Applications to Soft Matter}, Elsevier Science, 2013\relax
\mciteBstWouldAddEndPuncttrue
\mciteSetBstMidEndSepPunct{\mcitedefaultmidpunct}
{\mcitedefaultendpunct}{\mcitedefaultseppunct}\relax
\EndOfBibitem
\bibitem[Yesbolatova \emph{et~al.}(2022)Yesbolatova, Arai, Sakaue, and Kimura]{Sakaue_Kimura_PRL2022}
A.~K. Yesbolatova, R.~Arai, T.~Sakaue and A.~Kimura, \emph{Phys. Rev. Lett.}, 2022, \textbf{128}, 178101\relax
\mciteBstWouldAddEndPuncttrue
\mciteSetBstMidEndSepPunct{\mcitedefaultmidpunct}
{\mcitedefaultendpunct}{\mcitedefaultseppunct}\relax
\EndOfBibitem
\bibitem[Arai \emph{et~al.}(2017)Arai, Sugawara, Sato, Minakuchi, Toyoda, Nabeshima, Kimura, and Kimura]{Arai_2017}
R.~Arai, T.~Sugawara, Y.~Sato, Y.~Minakuchi, A.~Toyoda, K.~Nabeshima, H.~Kimura and A.~Kimura, \emph{Sce. Rep.}, 2017, \textbf{7}, 3631\relax
\mciteBstWouldAddEndPuncttrue
\mciteSetBstMidEndSepPunct{\mcitedefaultmidpunct}
{\mcitedefaultendpunct}{\mcitedefaultseppunct}\relax
\EndOfBibitem
\bibitem[Maeshima \emph{et~al.}(2024)Maeshima, Iida, Shimazoe, Tamura, and Ide]{Maeshima_2024}
K.~Maeshima, S.~Iida, M.~A. Shimazoe, S.~Tamura and S.~Ide, \emph{Trends in Cell Biology}, 2024, \textbf{34}, 7--17\relax
\mciteBstWouldAddEndPuncttrue
\mciteSetBstMidEndSepPunct{\mcitedefaultmidpunct}
{\mcitedefaultendpunct}{\mcitedefaultseppunct}\relax
\EndOfBibitem
\bibitem[Stukowski(2009)]{OVITO}
A.~Stukowski, \emph{Modelling and simulation in materials science and engineering}, 2009, \textbf{18}, 015012\relax
\mciteBstWouldAddEndPuncttrue
\mciteSetBstMidEndSepPunct{\mcitedefaultmidpunct}
{\mcitedefaultendpunct}{\mcitedefaultseppunct}\relax
\EndOfBibitem
\bibitem[Zidovska(2020)]{Zidovska2020}
A.~Zidovska, \emph{Biophysical Reviews}, 2020, \textbf{12}, 1093--1106\relax
\mciteBstWouldAddEndPuncttrue
\mciteSetBstMidEndSepPunct{\mcitedefaultmidpunct}
{\mcitedefaultendpunct}{\mcitedefaultseppunct}\relax
\EndOfBibitem
\bibitem[Put \emph{et~al.}(2019)Put, Sakaue, and Vanderzande]{Put2019}
S.~Put, T.~Sakaue and C.~Vanderzande, \emph{Phys. Rev. E}, 2019, \textbf{99}, 032421\relax
\mciteBstWouldAddEndPuncttrue
\mciteSetBstMidEndSepPunct{\mcitedefaultmidpunct}
{\mcitedefaultendpunct}{\mcitedefaultseppunct}\relax
\EndOfBibitem
\bibitem[Bruinsma \emph{et~al.}(2014)Bruinsma, Grosberg, Rabin, and Zidovska]{BRUINSMA2014}
R.~Bruinsma, A.~Grosberg, Y.~Rabin and A.~Zidovska, \emph{Biophysical Journal}, 2014, \textbf{106}, 1871--1881\relax
\mciteBstWouldAddEndPuncttrue
\mciteSetBstMidEndSepPunct{\mcitedefaultmidpunct}
{\mcitedefaultendpunct}{\mcitedefaultseppunct}\relax
\EndOfBibitem
\bibitem[Vandebroek and Vanderzande(2015)]{Vandebroek2015}
H.~Vandebroek and C.~Vanderzande, \emph{Phys. Rev. E}, 2015, \textbf{92}, 060601\relax
\mciteBstWouldAddEndPuncttrue
\mciteSetBstMidEndSepPunct{\mcitedefaultmidpunct}
{\mcitedefaultendpunct}{\mcitedefaultseppunct}\relax
\EndOfBibitem
\bibitem[Sakaue and Nakajima(2016)]{Sakaue_2016}
T.~Sakaue and C.~H. Nakajima, \emph{Phys. Rev. E}, 2016, \textbf{93}, 042502\relax
\mciteBstWouldAddEndPuncttrue
\mciteSetBstMidEndSepPunct{\mcitedefaultmidpunct}
{\mcitedefaultendpunct}{\mcitedefaultseppunct}\relax
\EndOfBibitem
\bibitem[Micheletti \emph{et~al.}(2024)Micheletti, Chubak, Orlandini, and Smrek]{Cristian_2024}
C.~Micheletti, I.~Chubak, E.~Orlandini and J.~Smrek, \emph{ACS Macro Letters}, 2024, \textbf{13}, 124--129\relax
\mciteBstWouldAddEndPuncttrue
\mciteSetBstMidEndSepPunct{\mcitedefaultmidpunct}
{\mcitedefaultendpunct}{\mcitedefaultseppunct}\relax
\EndOfBibitem
\bibitem[Chubak \emph{et~al.}(2022)Chubak, Pachong, Kremer, Likos, and Smrek]{Iurii_2022}
I.~Chubak, S.~M. Pachong, K.~Kremer, C.~N. Likos and J.~Smrek, \emph{Macromolecules}, 2022, \textbf{55}, 956--964\relax
\mciteBstWouldAddEndPuncttrue
\mciteSetBstMidEndSepPunct{\mcitedefaultmidpunct}
{\mcitedefaultendpunct}{\mcitedefaultseppunct}\relax
\EndOfBibitem
\bibitem[Awazu(2014)]{Akinori_2014}
A.~Awazu, \emph{Phys. Rev. E}, 2014, \textbf{90}, 042308\relax
\mciteBstWouldAddEndPuncttrue
\mciteSetBstMidEndSepPunct{\mcitedefaultmidpunct}
{\mcitedefaultendpunct}{\mcitedefaultseppunct}\relax
\EndOfBibitem
\bibitem[Halverson \emph{et~al.}(2014)Halverson, Smrek, Kremer, and Grosberg]{Halverson_2014}
J.~D. Halverson, J.~Smrek, K.~Kremer and A.~Y. Grosberg, \emph{Reports on Progress in Physics}, 2014, \textbf{77}, 022601\relax
\mciteBstWouldAddEndPuncttrue
\mciteSetBstMidEndSepPunct{\mcitedefaultmidpunct}
{\mcitedefaultendpunct}{\mcitedefaultseppunct}\relax
\EndOfBibitem
\bibitem[Rosa and Everaers(2008)]{Angelo_2008}
A.~Rosa and R.~Everaers, \emph{PLOS Computational Biology}, 2008, \textbf{4}, 1--10\relax
\mciteBstWouldAddEndPuncttrue
\mciteSetBstMidEndSepPunct{\mcitedefaultmidpunct}
{\mcitedefaultendpunct}{\mcitedefaultseppunct}\relax
\EndOfBibitem
\bibitem[Sakaue(2011)]{Sakaue_2011}
T.~Sakaue, \emph{Phys. Rev. Lett.}, 2011, \textbf{106}, 167802\relax
\mciteBstWouldAddEndPuncttrue
\mciteSetBstMidEndSepPunct{\mcitedefaultmidpunct}
{\mcitedefaultendpunct}{\mcitedefaultseppunct}\relax
\EndOfBibitem
\bibitem[Frank-Kamenetskii \emph{et~al.}(1975)Frank-Kamenetskii, Lukashin, and Vologodskii]{Frank-Kamenetskii1975}
M.~D. Frank-Kamenetskii, A.~V. Lukashin and A.~V. Vologodskii, \emph{Nature}, 1975, \textbf{258}, 398--402\relax
\mciteBstWouldAddEndPuncttrue
\mciteSetBstMidEndSepPunct{\mcitedefaultmidpunct}
{\mcitedefaultendpunct}{\mcitedefaultseppunct}\relax
\EndOfBibitem
\bibitem[Sakaue(2018)]{Sakaue_2018}
T.~Sakaue, \emph{Soft Matter}, 2018, \textbf{14}, 7507--7515\relax
\mciteBstWouldAddEndPuncttrue
\mciteSetBstMidEndSepPunct{\mcitedefaultmidpunct}
{\mcitedefaultendpunct}{\mcitedefaultseppunct}\relax
\EndOfBibitem
\bibitem[Shahamat and Rey(2015)]{SHAHAMAT2015100}
M.~Shahamat and A.~D. Rey, \emph{Chemical Engineering Science}, 2015, \textbf{121}, 100--109\relax
\mciteBstWouldAddEndPuncttrue
\mciteSetBstMidEndSepPunct{\mcitedefaultmidpunct}
{\mcitedefaultendpunct}{\mcitedefaultseppunct}\relax
\EndOfBibitem
\bibitem[Rasouli and Rey(2011)]{Rausouli2011}
G.~Rasouli and A.~D. Rey, \emph{The Journal of Chemical Physics}, 2011, \textbf{134}, 184901\relax
\mciteBstWouldAddEndPuncttrue
\mciteSetBstMidEndSepPunct{\mcitedefaultmidpunct}
{\mcitedefaultendpunct}{\mcitedefaultseppunct}\relax
\EndOfBibitem
\bibitem[Thompson \emph{et~al.}(2022)Thompson, Aktulga, Berger, Bolintineanu, Brown, Crozier, in~'t Veld, Kohlmeyer, Moore, Nguyen, Shan, Stevens, Tranchida, Trott, and Plimpton]{LAMMPS}
A.~P. Thompson, H.~M. Aktulga, R.~Berger, D.~S. Bolintineanu, W.~M. Brown, P.~S. Crozier, P.~J. in~'t Veld, A.~Kohlmeyer, S.~G. Moore, T.~D. Nguyen, R.~Shan, M.~J. Stevens, J.~Tranchida, C.~Trott and S.~J. Plimpton, \emph{Comp. Phys. Comm.}, 2022, \textbf{271}, 108171\relax
\mciteBstWouldAddEndPuncttrue
\mciteSetBstMidEndSepPunct{\mcitedefaultmidpunct}
{\mcitedefaultendpunct}{\mcitedefaultseppunct}\relax
\EndOfBibitem
\end{mcitethebibliography}
\bibliographystyle{rsc} 

\end{document}